\documentclass[12pt]{article} 

\usepackage{ol2}
\usepackage[draft]{hyperref}
\usepackage{amsmath}

\newcommand{\mum}{$\mu$m }
\newcommand{\mumd}{$\mu$m. }
\newcommand{\mumc}{$\mu$m, }

\begin{document}

\title{Towards visible CW pumped supercontinua}

\author{B. A. Cumberland,$^*$ J. C. Travers, S. V. Popov and J. R. Taylor}
\address{Femtosecond Optics Group, Imperial College London, London, SW7 2AZ, UK}
\address{$^*$Corresponding author: b.cumberland05@imperial.ac.uk}

\begin{abstract}We report a 1 \mum continuous wave pumped supercontinuum which extends short of the pump wavelength to 0.65 \mumd This is achieved by using a 50 W Yb fibre laser in combination with a photonic crystal fibre with a carefully engineered zero dispersion wavelength. We show that the short wavelength generation is due to a combination of four-wave mixing and dispersive wave trapping by solitons. The evolution and limiting factors of the continuum are discussed.\end{abstract}

Research on supercontinuum generation was reinvigorated after the advent of photonic crystal fibres (PCF), owing to the high non-linearity and controllable dispersion possible in these fibres\cite{Knight:1996p84}. A wide variety of pump sources and PCFs have been used including fs, ps, ns and continuous wave (CW) pumps\cite{Dudley:2006p154} along with low water loss \cite{Travers:2005p445}, long tapers\cite{Kudlinski:2006p167} and high IR transmission PCFs\cite{Stone:2008p436} to optimise the continuum output in various ways. CW pumped supercontinua offer several advantages over pulse pumped continua including higher spectral power densities and smoother spectra. They are therefore of great interest for certain applications and have been demonstrated in highly nonlinear fibre and PCF pumped at 1.5 \mum and 1.06 \mum respectively\cite{Prabhu:2000p538,Avdokhin:2003p150}. While the generation of wavelengths short of the pump has been seen at 1.5 \mum \cite{Dudley:2006p154}, short wavelength generation has not been observed with CW pumping at 1 \mumd The prospect of high spectral power densities similar to that reported in\cite{Cumberland:2008p31} in the visible would open up a variety of new applications for supercontinuum sources. 

In this paper we present, for the first time to our knowledge, experimental results showing a 1.07 \mum CW laser pumped supercontinuum extending to wavelengths short of the pump.  We attribute the short wavelength generation to FWM and the trapping of dispersive waves by solitons (here after called soliton trapping) and examine the limiting factors. In order to initiate either of these processes the pump must be very close to the zero dispersion wavelength (ZDW) in the anomalous regime. This is so that energy can be transferred to the short wavelength side in the normal dispersion regime. We believe that this tight tolerance on the location of the pump with respect to the ZDW is the reason that short wavelength generation has not been observed at 1 \mum before.

Depending upon the pump regime there are several different mechanisms involved in generating wavelengths short of the pump. In the case of pumping with a fs source in the anomalous regime, the fs pulses can be considered high order solitons in the fibre. These rapidly broaden and then fission to fundamental solitons. During the fission process excess energy is shed as dispersive waves on the short wavelength side. Generally these dispersive waves undergo no further shifting\cite{Dudley:2006p154}. Under certain circumstances, it is possible that these dispersive waves can be coupled to the solitons\cite{Gorbach:2007p299,Beaud:1987p455}. This means that as the soliton self-frequency shifts (SSFS) to longer wavelengths the coupled dispersive wave is shifted to shorter wavelengths as dictated by the group velocity matching conditions. This soliton trapping effect has also been demonstrated with a ps pump source\cite{Stone:2008p436}. In the ps and ns regimes phase matched FWM can make a significant contribution to short wavelength generation. The dispersion curve of the fibre determines a set of phase matching conditions leading to degenerate and non-degenerate FWM\cite{Dudley:2006p154}.

In the CW pump regime the supercontinuum evolves from noise. Typically the CW field is broken up by MI which gives rise to first order solitons. These solitons then undergo intra-pulse Raman scattering leading to the SSFS, generating a long wavelength soliton Raman continuum. Clearly, as no high order solitons are generated, the fission process described above is not a feasible mechanism. This leaves FWM and soliton trapping as possible mechanisms to utilise for short wavelength generation. Soliton trapping requires a dispersive wave in the normal dispersion region which can be trapped. In order for the trapping mechanism to work the soliton has to have a sufficiently short duration to generate the required index modulation to trap the dispersive radiation. Once trapped, this energy can be shifted to shorter wavelengths as the soliton self-frequency shifts to longer wavelengths, provided the group velocity matching conditions are met\cite{Gorbach:2007p299}. In the CW regime, the initial solitons generated from the MI process have to be short enough (in time) so that spectrally, part of the soliton overlaps into the normal dispersion regime to generate the required dispersive wave. For FWM the phase matching conditions must be met. Experimental experience suggests, that with low pump powers, FWM generally requires some seeding. One way to achieve this, is to have the anti-Stokes MI sideband extend into the normal dispersion regime. Thus both these mechanisms, place very tight constraints on the ZDW relative to the pump, given the low pump powers involved. With this knowledge it is possible to choose a PCF with suitable parameters to demonstrate these effects.

The experimental set-up simply consisted of a CW pump laser and a PCF spliced to it. The pump laser was a 50 W CW Yb fibre laser (IPG Photonics) with a single-mode output fibre. The lasers emission was centred on 1.071 \mum with a linewidth of 0.3 nm. This was directly spliced to a 100 m long PCF (Crystal Fiber) with a zero dispersion wavelength of 1.068 \mumc nonlinear coefficient of 9.8 (W km)$^{-1}$ and mode field diameter of 5 \mumd The loss was measured as being 9 dB/km with a 130 dB/km peak at the OH$^-$ induced loss at 1.38 \mumd The splice loss was 0.6-0.85 dB and was thermally managed at high pump powers. The output end of the PCF was angle cleaved to reduce back reflections which can seed strong Raman-Stokes lines in the continuum. The fibre was pumped in two regimes, CW and modulated. Under the modulated regime the peak power was enhanced by a factor of $\sim$2.5. The spectra were measured on an Advantest (Q8384) and visible Anritsu (MS9030A/MS9701B) optical spectrum analyser (OSA).

\begin{figure}[htb]
\centerline{\includegraphics[width=8.0cm]{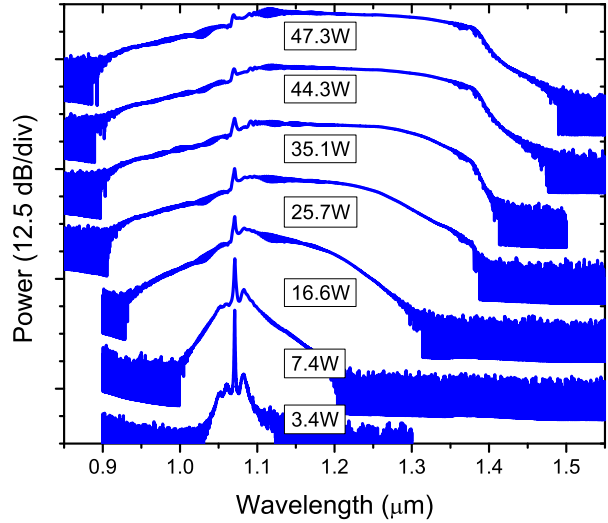}}
\caption{Evolution of the supercontinuum with increasing pump power. Powers marked represent the power launched after the splice. For 47.3 W the continuum output power was 32.6 W. }
\label{fig:evolution}
\end{figure}

Fig. \ref{fig:evolution} shows the evolution of the supercontinuum under CW pumping for several different pump powers. The MI side bands are clearly visible at low pump powers with a continuous extensions to longer and shorter wavelengths as the pump power is increased. Eventually the long wavelength edge of the continuum is limited by OH$^-$ loss at 1.38 \mumd The anti-Stokes side is 5 dB down compared with the Stokes side immediately next to the pump. This may be due to limited energy transfer to the short wavelength side initially, combined with losses due to the short wavelength side pumping the long wavelength side via Raman. It should also be noted that the noise floor of the OSA is 17 dB higher around 0.9 \mumc compared with that above 1.0 \mumd The spectral evolution of this continuum can be explained by examining the phase matching curves shown in Fig. 2.

\begin{figure}[htb]
\centerline{\includegraphics[width=8.0cm]{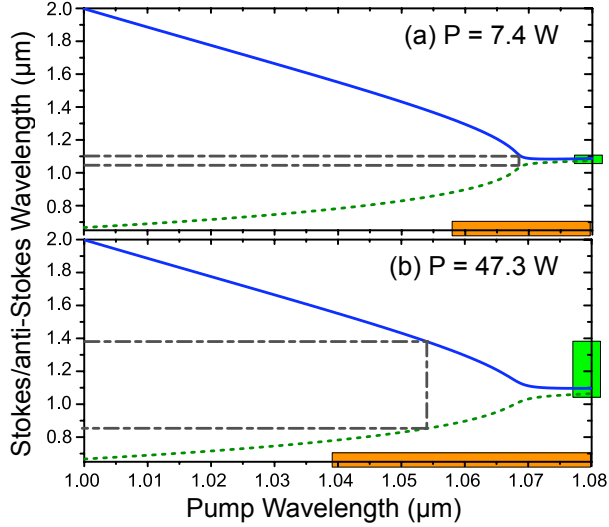}}
\caption{Phase matching curves for FWM.}
\end{figure}

Fig. 2 shows the phase matching curves for pump powers of 7.4 W (a) and 47.3 W (b). The Stokes and anti-Stokes curves are solid (blue) and dotted (green) respectively. The available short pump wavelengths due to MI are denoted by the large shaded (orange) bar along the bottom $x$ axis. The available long seed wavelengths due to MI or the soliton Raman continuum are marked on the right $y$ axis by the large shaded (green) bar. The upper and lower wavelength limits available via FWM are denoted by the dashed (grey) lines intersecting the phase matching curve at the limiting seed wavelength. These phase matching curves were calculated for the 12 different pump powers and found to be in good agreement with the spectral evolution in every case. The most likely FWM process is $\omega_{pump}+\omega_{MI_{short}}=\omega_{Stokes}+\omega_{anti-Stokes}$. Notably the Stokes appears to be seeded by the soliton Raman continuum and is the limiting factor on the anti-Stokes wavelength. One would expect, that as the pump wavelength is shifted further from the ZDW, there will come a point when the short wavelengths generated via MI become the limiting factor, on the anti-Stokes wavelength. Unfortunately, it is impossible to rule out soliton trapping as a mechanism in this case. As can be seen from Fig. 4, the matching conditions are almost identical until 1.3 \mumd We currently favour FWM as an explanation because the phase matched curve is in slightly better agreement with the data at the longer wavelengths. An experiment involving shifting the pump wavelength combined with numerical simulations should help clarify the dynamics of this process. Finally it is clear that if we want to extend to much shorter wavelengths, the long wavelength edge of the continuum needs to extend beyond the OH$^-$ induced loss at 1.38 \mumd To demonstrate this, we modulated the pump to produce $\mu$s pulses. The results are shown in Fig. 3.

\begin{figure}[htb]
\centerline{\includegraphics[width=8.0cm]{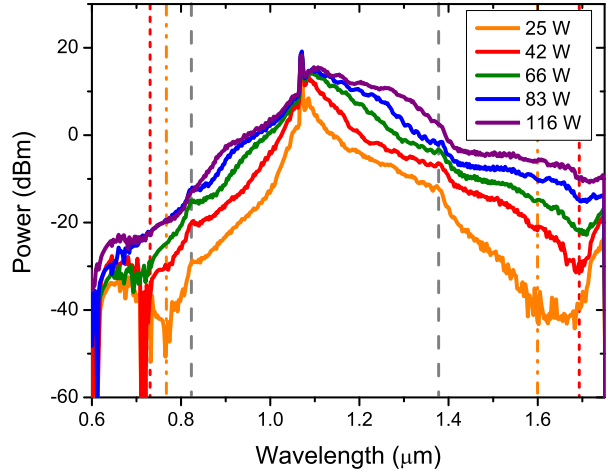}}
\caption{Supercontinuum evolution with pump power using a modulated pump source. Dashed lines mark particular spectral features.}
\label{fig:modspec}
\end{figure}

Modulating the CW laser enhances the continuum beyond that expected from the peak power increase alone. Measurements suggest that the noise dynamics change, enhancing the MI process. Regardless of the ehancement, Fig. 3 shows that a number of solitons now have enough energy to cross the OH$^-$ loss at 1.38 \mumc resulting in short wavelength generation down to 650 nm. Analysing the continuum evolution shows that the short wavelength edge appears to expand faster than that allowed by FWM alone. To explain this increased rate of expansion, one has to invoke soliton trapping as a mechanism. The dashed lines in figure 3 mark corresponding spectral features. The dip due to the OH$^-$ loss at 1.38 \mum has a corresponding spectral feature at 0.82 \mumd Examining this feature on the group velocity and phase matching curves, shown in Fig. 4, it is clear that group velocity matching, thus soliton trapping, is the better fit. However, spectral features at longer wavelengths (denoted by the orange and red dashed lines) are in much better agreement with the FWM phase matched curve. Hence it seems likely that both FWM and soliton trapping are playing an active role in the short wavelength generation. One may speculate that the OH$^-$ loss makes the solitons increase in duration thus reducing their power, consequently the intensity dependent refractive index reduces so the cross-phase modulation effect reduces, weakening the trapping and possibly allowing FWM to dominate. Alternatively the loss may simply act as a barrier reducing the number of solitons shifting to longer wavelengths and their corresponding trapped dispersive waves\cite{Cumberland:2008p31}. In either case, FWM appears to become more significant at the shorter wavelengths. The short wavelength edge of the continuum is now defined by the high losses in the PCF beyond 2 \mumd 
  
\begin{figure}[htb]
\centerline{\includegraphics[width=8.0cm]{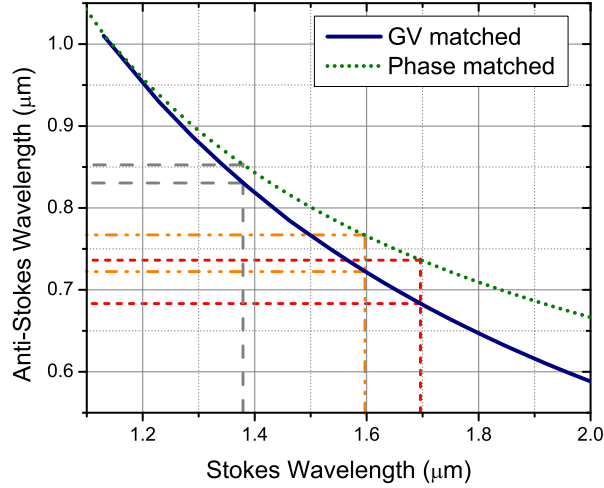}}
\caption{Group velocity and phase matched curves for our PCF. Corresponding spectral features of Fig. \ref{fig:modspec} are marked.}
\end{figure}

In conclusion we have shown the first short wavelength generation in a 1 \mum CW pumped supercontinuum. The short wavelength generation has been shown to be due to a combination of FWM and soliton trapping. Improvements should be possible by a variety of mechanisms including pumping at much higher powers \cite{travers08}, optimisation of the PCF to aid soliton trapping\cite{Stone:2008p436}, dispersion management via tapers\cite{Kudlinski:2006p167} as well as numerous schemes to enhance the gain and energy on the short wavelength side. Although these initial results are promising, it is likely that much work remains before CW pumped supercontinuum sources with high spectral power densities across the visible will be realised.

\bibliographystyle{ol}

\begin{thebibliography}{10}
\newcommand{\enquote}[1]{``#1''}

\bibitem{Knight:1996p84}
J.~C. Knight, T.~Birks, P.~Russell, and D.~Atkin, Opt. Lett. \textbf{21}, 1547
  (1996).

\bibitem{Dudley:2006p154}
J.~Dudley, G.~Genty, and S.~Coen, Rev. Mod. Phys. \textbf{78}, 1135 (2006).

\bibitem{Travers:2005p445}
J.~C. Travers, R.~Kennedy, S.~V. Popov, J.~R. Taylor, H.~Sabert, and B.~Mangan,
  Opt. Lett. \textbf{30}, 1938 (2005).

\bibitem{Kudlinski:2006p167}
A.~Kudlinski, A.~K. George, J.~C. Knight, J.~C. Travers, A.~Rulkov, S.~V.
  Popov, and J.~R. Taylor, Opt. Express \textbf{14}, 5715 (2006).

\bibitem{Stone:2008p436}
J.~M. Stone and J.~C. Knight, Opt. Express \textbf{16}, 2670 (2008).

\bibitem{Prabhu:2000p538}
M.~Prabhu, N.~Kim, and K.~Ueda, Jpn J Appl Phys 2 \textbf{39}, L291 (2000).

\bibitem{Avdokhin:2003p150}
A.~Avdokhin, S.~V. Popov, and J.~R. Taylor, Opt. Lett. \textbf{28}, 1353
  (2003).

\bibitem{Cumberland:2008p31}
B.~A. Cumberland, J.~C. Travers, S.~V. Popov, and J.~R. Taylor, Opt. Express
  \textbf{16}, 5954 (2008).

\bibitem{Gorbach:2007p299}
A.~Gorbach and D.~Skryabin, Phys. Rev. A \textbf{76}, 053803 (2007).

\bibitem{Beaud:1987p455}
P.~Beaud, W.~Hodel, B.~Zysset, and H.~Weber, Ieee J Quantum Elect \textbf{23},
  1938 (1987).

\bibitem{travers08}
J.~C. Travers, A.~B. Rulkov, S.~V. Popov, and J.~R. Taylor, \enquote{Visible
  supercontinuum generation in photonic crystal fibre with a 400 w continuous
  wave fibre laser}, In preparation.

\end{thebibliography}

\end{document}